\documentclass[10pt]{iopart}
\usepackage{graphicx}
\usepackage{iopams}
\usepackage{cite}

\def\prl {Phys.\ Rev.\ Lett.\ }
\def\pra {Phys.\ Rev.\ A}
\def\prb {Phys.\ Rev.\ B}
\def\pre {Phys.\ Rev.\ E}

\def\figwidth{4.65cm}

\begin{document}

\paper[Smeared phase transition in a disordered contact process]
{Monte-Carlo simulations of the smeared phase transition in a contact process with
extended defects}

\author{Mark Dickison and Thomas Vojta}

\address{Department of Physics, University of Missouri - Rolla, Rolla, MO 65409, USA}

\begin{abstract}
We study the nonequilibrium phase transition in a contact process with extended quenched
defects by means of Monte-Carlo simulations. We find that the spatial disorder
correlations dramatically increase the effects of the impurities. As a result, the sharp
phase transition is completely destroyed by smearing. This is caused by effects similar
to but stronger than the usual Griffiths phenomena, viz., rare strongly coupled spatial
regions can undergo the phase transition independently from the bulk system.  We
determine both the stationary density in the vicinity of the smeared transition and its
time evolution, and we compare the simulation results to a recent theory based on
extremal statistics.

\end{abstract}


\submitto{\JPA}
%\maketitle

%%%%%%%%%%%%%%%%%%%%%%%%%%%%%%%%%%%%%%%%%%%%%%%%%%%%%%%%%%%%%%%%%%%%%%%%%%%%%%%%%%%%%%%%%%%%
\section{Introduction}
\label{sec:intro}
%%%%%%%%%%%%%%%%%%%%%%%%%%%%%%%%%%%%%%%%%%%%%%%%%%%%%%%%%%%%%%%%%%%%%%%%%%%%%%%%%%%%%%%%%%%%

Rare regions are an important, if intricate, aspect of systems with impurities and
defects. In recent years, their influence on phase transitions and critical phenomena has
reattracted considerable attention. Rare region effects were first studied in the context
of classical equilibrium phase transitions. Griffiths \cite{Griffiths} showed that they
lead to a singular free energy in an entire parameter region in the vicinity of the phase
transition, now known as the Griffiths region. However, in classical systems with
uncorrelated disorder, this Griffiths singularity in the free energy is an essential one
and thus very weak and probably unobservable in experiment. Disorder correlations
generically increase the effects of impurities. Therefore, stronger rare region effects
have been found in classical systems with extended defects and in random quantum systems
(where the correlations are in imaginary time direction). In the random transverse field
Ising model \cite{dsf9295}, or equivalently, the classical McCoy-Wu model \cite{McCoyWu},
the Griffiths singularity takes a power law form, accompanied by a diverging magnetic
susceptibility in the Griffiths region. Very recently, it has been found that some phase
transitions can even be completely destroyed by smearing when the rare regions order
independently from the bulk system. This happens, e.g.,  in a classical Ising magnet with
planar defects \cite{us_planar} and in itinerant quantum ferromagnets \cite{us_rounding}.

In this paper, we investigate the effects of rare regions on \emph{nonequilibrium} phase
transitions with quenched spatial disorder. We concentrate on the prominent class of
absorbing state phase transitions which separate active, fluctuating states from
inactive, absorbing states where fluctuations cease entirely
\cite{chopard_book,marro_book,hinrichsen00,tauber}. The generic universality class for
absorbing state transitions is directed percolation (DP) \cite{dp}. According to a
conjecture by Janssen and Grassberger \cite{conjecture}, all absorbing state transitions
with a scalar order parameter, short-range interactions, and no extra symmetries or
conservation laws belong to this class. Examples include the transitions in the contact
process \cite{contact}, catalytic reactions \cite{ziff}, interface growth \cite{tang}, or
turbulence \cite{turb}.

The effects of \emph{uncorrelated} spatial disorder, i.e., point-like defects, on the DP
transition have been studied in some detail in the past. According to the Harris
criterion \cite{harris,noest}, the DP universality class is unstable against spatial
disorder, because the (spatial) correlation length exponent $\nu_\perp$ violates the
inequality $\nu_\perp > 2/d$ for all spatial dimensionalities $d<4$. Indeed, in the
corresponding field theory, spatial disorder leads to runaway flow of the renormalization
group (RG) equations \cite{janssen97}, destroying the DP behavior. Several other studies
\cite{bramson,moreira,webman,cafiero} agreed on the instability of DP against spatial
disorder, but a consistent picture has been slow to evolve. Recently, Hooyberghs {\it et
al.} applied the Hamiltonian formalism \cite{alcaraz} to the contact process with spatial
disorder \cite{hooyberghs}. Using a version of the Ma-Dasgupta-Hu strong-disorder RG
\cite{SDRG} these authors showed that the transition (at least for sufficiently strong
disorder) is controlled by an exotic infinite-randomness fixed point with activated
rather than the usual power-law scaling.

Very recently, it has been suggested \cite{contact_pre} that extended spatial defects
like dislocations, disordered layers, or grain boundaries can have an even more dramatic
effect on nonequilibrium phase transitions in the DP universality class. Rare region
effects similar to but stronger than the usual Griffiths phenomena \cite{Griffiths,noest}
actually destroy the sharp transition by smearing. This happens because rare strongly
coupled spatial regions can undergo the transition independently from the bulk system.
Based on an extremal statistics approach it has been predicted \cite{contact_pre} that
the spatial density distribution in the tail of the smeared transition is very
inhomogeneous, with the average stationary density and the survival probability depending
exponentially on the control parameter.

In the present paper we present results of extensive Monte-Carlo simulations of a
two-dimensional contact process with linear spatial defects which provide numerical
evidence for this smearing scenario in a realistic model with short-range couplings. The
paper is organized as follows.  In section \ref{sec:theory}, we introduce the model and
briefly summarize the results of the extremal statistics theory for the smeared phase
transition. In section \ref{sec:montecarlo} we present our simulation method and the
numerical results together with a comparison to the theoretical predictions. We conclude
in section \ref{sec:conclusions} by discussing the importance of our results and their
generality.

%%%%%%%%%%%%%%%%%%%%%%%%%%%%%%%%%%%%%%%%%%%%%%%%%%%%%%%%%%%%%%%%%%%%%%%%%%%%%%%%%%%%%%%%%%%%
\section{Theory}
\label{sec:theory}
%%%%%%%%%%%%%%%%%%%%%%%%%%%%%%%%%%%%%%%%%%%%%%%%%%%%%%%%%%%%%%%%%%%%%%%%%%%%%%%%%%%%%%%%%%%%

\subsection{Contact process with extended impurities}

The contact process \cite{contact} is a prototypical system in the directed percolation
universality class. It can be interpreted, e.g., as a model for the spreading of a
disease. The contact process is defined on a $d$-dimensional hypercubic lattice. Each
lattice site $\mathbf r$ can be active (occupied by a particle) or inactive (empty).
During the time evolution, active sites can infect their neighbors or they can
spontaneously become inactive. Specifically, particles are created at empty sites at a
rate $\lambda n/ (2d)$ where $n$ is the number of active nearest neighbor sites and the
`birth rate' $\lambda$ is the control parameter. Particles are annihilated at unit rate.
For small birth rate $\lambda$, annihilation dominates, and the absorbing state without
any particles is the only steady state (inactive phase). For large birth rate $\lambda$,
there is a steady state with finite particle density (active phase).  The two phases are
separated by a nonequilibrium phase transition in the DP universality class at
$\lambda=\lambda_c^0$.

Quenched spatial disorder can be introduced by making the birth rate $\lambda$ a random
function of the lattice site. Point-like defects are described by spatially uncorrelated
disorder. We are interested in the case of extended defects which can be described by
disorder perfectly correlated in $d_{cor}$ dimensions, but uncorrelated in the remaining
$d_r=d-d_{cor}$ dimensions. Here $d_{cor}=1$ and 2 corresponds to linear and planar
defects, respectively. Thus, $\lambda$ is a function of ${\mathbf r}_r$ which is the
projection of the position vector $\mathbf r$ on the uncorrelated directions. For
definiteness, we assume that the birthrate values $\lambda({\mathbf r}_r)$ are drawn from
a binary probability distribution
\begin{equation}
P[\lambda({\mathbf r}_r)] = (1-p)\, \delta[\lambda({\mathbf r}_r)-\lambda] + p\,
\delta[\lambda({\mathbf r}_r) - c\lambda]
\end{equation}
where $p$ and $c$ are constants between 0 and 1. In other words, there are extended
impurities of spatial density $p$ where the birth rate $\lambda$ is reduced by a factor
$c$.

\subsection{Smeared phase transition}

In this subsection, we briefly summarize the arguments leading to the smearing of the
phase transition and the predictions of the extremal statistics theory \cite{contact_pre}
to the extent necessary for the comparison with the Monte-Carlo results.

In analogy to the Griffiths phenomena \cite{Griffiths,noest}, there is a small but finite
probability $w$ for finding a large spatial region of linear size $L_r$ (in the
uncorrelated directions) devoid of impurities. Up to pre-exponential factors, this
probability is given by
\begin{equation}
w \sim \exp(-\tilde p L_r^{d_r}) \label{eq:rr}
\end{equation}
with  $\tilde p = -\ln(1-p)$. These rare regions can be locally in the active phase, even
if the bulk system is still in the inactive phase. Since the impurities in our system are
extended, each rare region is infinite in $d_{cor}$ dimensions but finite in the
remaining $d_r$ dimensions. This is a crucial difference to systems with uncorrelated
disorder, where the rare regions are finite. In our system, each rare region can
therefore undergo a true phase transition {\em independently} of the rest of the system
at some $\lambda_c(L_r)>\lambda_c^0$. According to finite-size scaling \cite{barber},
\begin{equation}
\lambda_c(L_r) - \lambda_c^0 = A L_r^{-\phi}~, \label{eq:shift}
\end{equation}
where $\phi$ is the clean ($d$-dimensional) finite-size scaling shift exponent and $A$ is
the amplitude for the crossover from a $d$-dimensional bulk system to a `slab' infinite
in $d_{cor}$ dimensions but but finite in $d_r$ dimensions. If the total dimensionality
$d=d_{cor}+d_r<4$, hyperscaling is valid, and $\phi = 1/\nu_\bot$ which we assume from
now on.

The resulting global phase transition is very different from a conventional continuous
phase transition, where a nonzero order parameter develops as a collective effect of the
entire system, accompanied by a diverging correlation length in all directions. In
contrast, in our system, the order parameter develops very inhomogeneously in space with
different parts of the system (i.e., different ${\bf r}_r$ regions) ordering
independently at different $\lambda$. Correspondingly, the correlation length in the
uncorrelated directions remains finite across the transition. This defines a smeared
transition.

In order to determine the global system properties in the vicinity of the smeared
transition, we  combine (\ref{eq:rr}) and (\ref{eq:shift}) to obtain the probability for
finding a rare region which becomes active at $\lambda_c$ as
\begin{equation}
w(\lambda_c) \sim \exp \left( -B (\lambda_c -\lambda_c^0)^{-d_r\nu_\bot} \right)
\label{eq:w_lam}
\end{equation}
for $\lambda_c-\lambda_c^0 \to 0+$. Here, $B= \tilde p A^{d_r\nu_\bot}$.

The total density $\rho$ (the total number of active sites) at a certain $\lambda$ is
obtained by summing over all active rare regions, i.e., all regions with $\lambda >
\lambda_c$. Since the functional dependence on $\lambda$ of the density on any given
active island is of power-law type it does not enter the leading exponentials but only
the pre-exponential factors. Thus, the stationary density develops an exponential tail,
\begin{equation}
\rho_{st}(\lambda) \sim \exp \left( -B (\lambda-\lambda_c^0)^{-d_r\nu_{\bot}} \right)~,
\label{eq:rho}
\end{equation}
for all birth rates $\lambda$ above the clean critical point $\lambda_c^0$. Analogous
arguments can be made for the survival probability $P(\lambda$) of a single seed site. If
the seed site is on an active rare region it will survive with a probability that depends
on $\lambda$ via a power law. If it is not on an active rare region, the seed will die.
To exponential accuracy the survival probability is thus also given by (\ref{eq:rho}).
The local spatial density distribution in the tail of the smeared transition is very
inhomogeneous. On active rare regions, the density is of the same order of magnitude as
in the clean system. Away from these regions it is exponentially small.

We now turn to the dynamics in the tail of the smeared transition. The long-time decay of
the density (starting from a state with $\rho=1$) is dominated by the rare regions while
the bulk contribution decays exponentially. According to finite size scaling
\cite{barber}, the behavior of the correlation time $\xi_t$ of a single rare region of
size $L_r$ in the vicinity of the clean bulk critical point can be modelled by
\begin{equation}
\xi_t(\Delta, L_r) \sim L_r^{(z\nu_\bot - \tilde z\tilde \nu_\bot)/\nu_\bot}
   \left|  \Delta - A L_r ^{-1/\nu_\bot}\right|^{-\tilde z \tilde \nu_\bot}~. \label{eq:xit}
\end{equation}
Here $\Delta=\lambda-\lambda_c^0>0$, $z$ is the $d$-dimensional bulk dynamical critical
exponent, and $\tilde \nu_\bot$ and $\tilde z$ are the correlation length and dynamical
exponents of a $d_r$-dimensional system. To exponential accuracy, the time dependence of
the total density is given by
\begin{equation}
\rho(\lambda,t) \sim \int dL_r ~\exp \left [-\tilde p L_r^{d_r} -D t/\xi_t(\Delta,L_r)
\right]\label{eq:sp_int}
\end{equation}
where $D$ is a constant.

Let us first consider the time evolution at the clean critical point
$\lambda=\lambda_c^0$. For $\Delta=0$, the correlation time (\ref{eq:xit}) simplifies to
$\xi_t \sim L_r^z$.  Using the saddle point method to evaluate the integral
(\ref{eq:sp_int}), we find the leading long-time decay of the density to be given by a
stretched exponential,
\begin{equation}
\ln \rho(t) \sim - \tilde{p}^{z/(d_r+z)}~ t^{d_r/(d_r+z)}~. \label{eq:stretched}
\end{equation}

For $\lambda<\lambda_c^0$, i.e, in the absorbing phase, the correlation time of the
largest islands does not diverge but is cut-off by the distance from the clean critical
point, $\xi_t \sim \Delta^{-z\nu}$. The large islands with this correlation time dominate
the variational integral (\ref{eq:sp_int}). This leads to a simple exponential decay with
a decay constant $\tau \sim \Delta^{-z\nu}$

The most interesting case is $\lambda>\lambda_c^0$, i.e., the tail region of the smeared
transition. Here, we repeat the saddle point analysis with the full expression
(\ref{eq:xit}) for the correlation time. For intermediate times $t<t_x\sim
(\lambda-\lambda_c^0)^{-(d_r+z)\nu_\bot}$ the decay of the average density is still given
by the stretched exponential (\ref{eq:stretched}). For times larger than the crossover
time $t_x$ the system realizes that some of the rare regions are in the active phase and
contribute to a finite steady state density. The approach of the average density to this
steady state value is characterized by a power-law.
\begin{equation}
\rho(t) - \rho(\infty) \sim t^{-\psi}~. \label{eq:power}
\end{equation}
The value of $\psi$ cannot be found by our methods since it depends on the neglected
nonuniversal pre-exponential factors.

%%%%%%%%%%%%%%%%%%%%%%%%%%%%%%%%%%%%%%%%%%%%%%%%%%%%%%%%%%%%%%%%%%%%%%%%%%%%%%%%%%%%%%%%%%%%
\section{Monte-Carlo simulations}
\label{sec:montecarlo}
%%%%%%%%%%%%%%%%%%%%%%%%%%%%%%%%%%%%%%%%%%%%%%%%%%%%%%%%%%%%%%%%%%%%%%%%%%%%%%%%%%%%%%%%%%%%

\subsection{Simulation method}

We now illustrate the smearing of the phase transition by extensive computer simulation
results for a 2d contact process with linear defects ($d_{cor}=d_r=1$). There is a number
of different ways to actually implement the contact process on the computer (all
equivalent with respect to the universal behavior). We follow the widely used algorithm
described, e.g., by Dickman \cite{dickman99}. Runs start at time $t=0$ from some
configuration of occupied and empty sites. Each event consists of randomly selecting an
occupied site $\mathbf{r}$ from a list of all $N_p$ occupied sites, selecting a process:
creation with probability $\lambda(\mathbf{r}_r)/[1+ \lambda(\mathbf{r}_r)]$ or
annihilation with probability $1/[1+ \lambda(\mathbf{r}_r)]$ and, for creation, selecting
one of the neighboring sites of $\mathbf{r}$. The creation succeeds, if this neighbor is
empty. The time increment associated with this event is $1/N_p$.

Using this algorithm, we have performed simulations for linear system sizes up to
$L=3000$ and impurity concentrations $p=0.2,0.25,0.3,0.35$ and 0.4. The relative strength
of the birth rate on the impurities was $c=0.2$ for all simulations. The data presented
below represent averages of $200$ disorder realizations. Because of the large system
sizes and the high number of realizations, the statistical errors are very low. The
\emph{relative} statistical error of the average density ranges from
$\delta\rho/\rho\approx 10^{-4}$ at high densities to $\delta\rho/\rho\approx 10^{-2}$
for $\rho\approx 10^{-3}$ and to $\delta\rho/\rho\approx 10^{-1}$ for the lowest shown
densities $\rho \approx 10^{-6}$. Except at these lowest densities, the error is smaller
than or comparable with the line width in the figures.

We have chosen the above parameters, viz., a low concentration of impurities which have a
birth rate much smaller than the bulk, because these conditions are favorable for
observing the smeared transition in a finite size simulation. If $p$ was too large, the
exponential drop-offs in eqs.\ (\ref{eq:rho}) and (\ref{eq:stretched}) would be very
steep and hard to observe over a significant range of $\lambda$ or $t$, respectively. If
$c$ was too close to one, clean critical fluctuations would mask the tail of the smeared
transition.

\subsection{Time evolution}

\begin{figure}
\includegraphics[width=\figwidth,angle=-90]{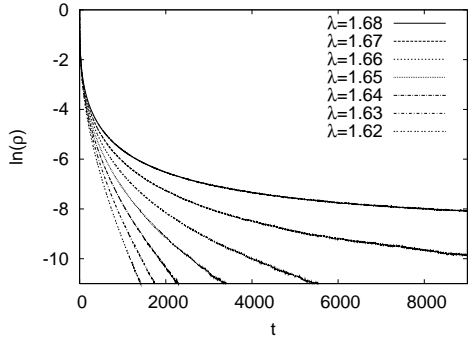}
\caption{Overview of the time evolution of the density $\rho$ for a system with $L=3000$
and $p=0.2$ and several birth rates ($\lambda=1.68,\ldots,1.62$ from top to bottom) in
the vicinity of the clean critical point $\lambda_c^0=1.6488$.} \label{fig:evo}
\end{figure}

In this subsection, we discuss the time evolution of the density starting from a
completely occupied lattice, $\rho(0)=1$. Figure \ref{fig:evo} presents an overview of
the behavior of a system with impurity concentration $p=0.2$, system size $L=3000$, and
several birth rates from $\lambda=1.62 \ldots 1.68$.
The clean critical point is at $\lambda_c^0=1.6488$ \cite{moreira}. The figure shows that
the long-time decay of the density in the absorbing phase, $\lambda<\lambda_c^0$, is
approximately exponential, in agreement with the expectation discussed after eq.\
(\ref{eq:stretched}) The decay constant of this exponential increases with decreasing
$\lambda$. In contrast, for $\lambda>\lambda_c^0$ the density approaches a nonzero value
in the long-time limit. Close to $\lambda_c^0$, the density appears to decay, but slower
than exponentially.

\begin{figure}
\includegraphics[width=\figwidth,angle=-90]{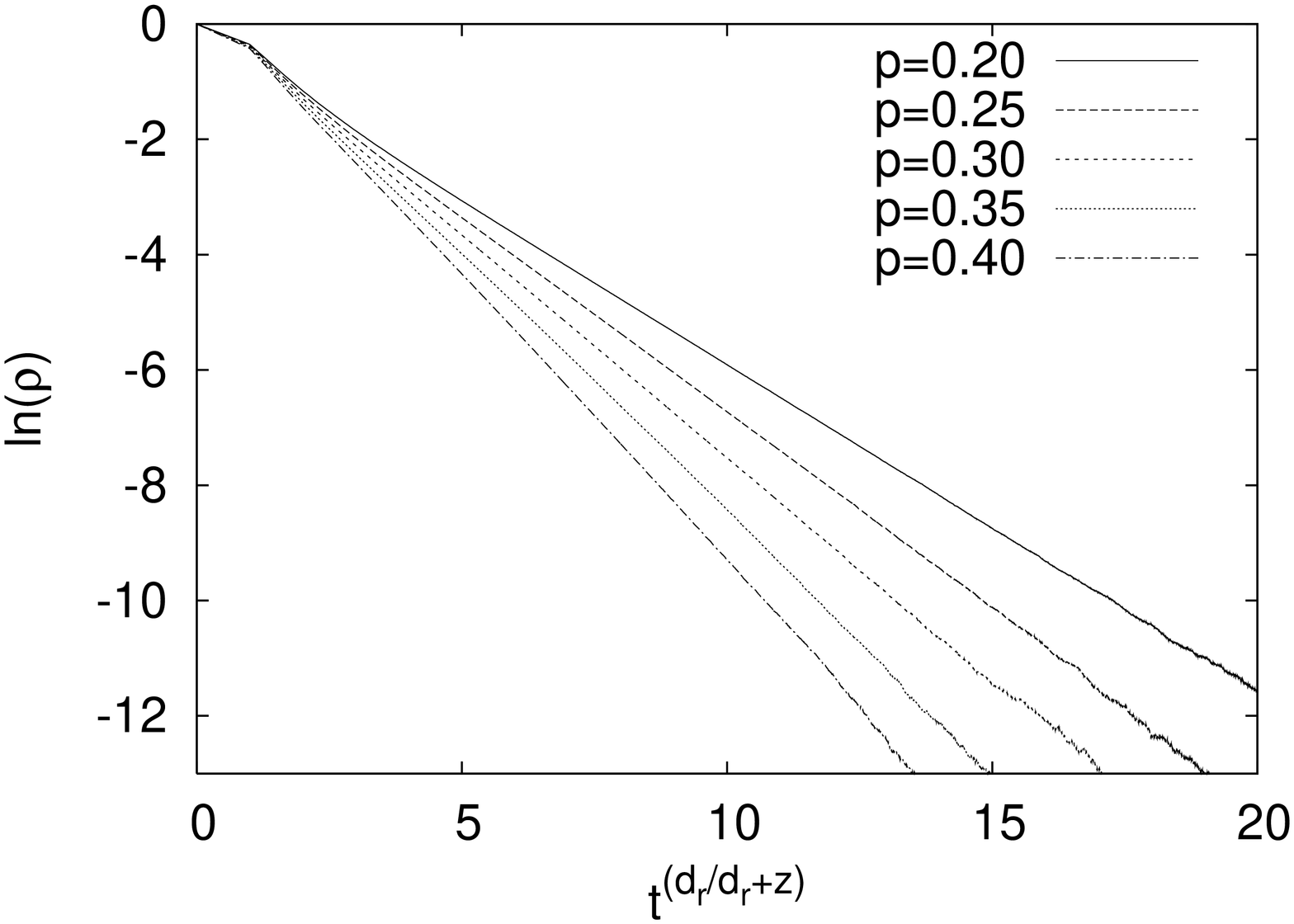}\includegraphics[width=\figwidth,angle=-90]{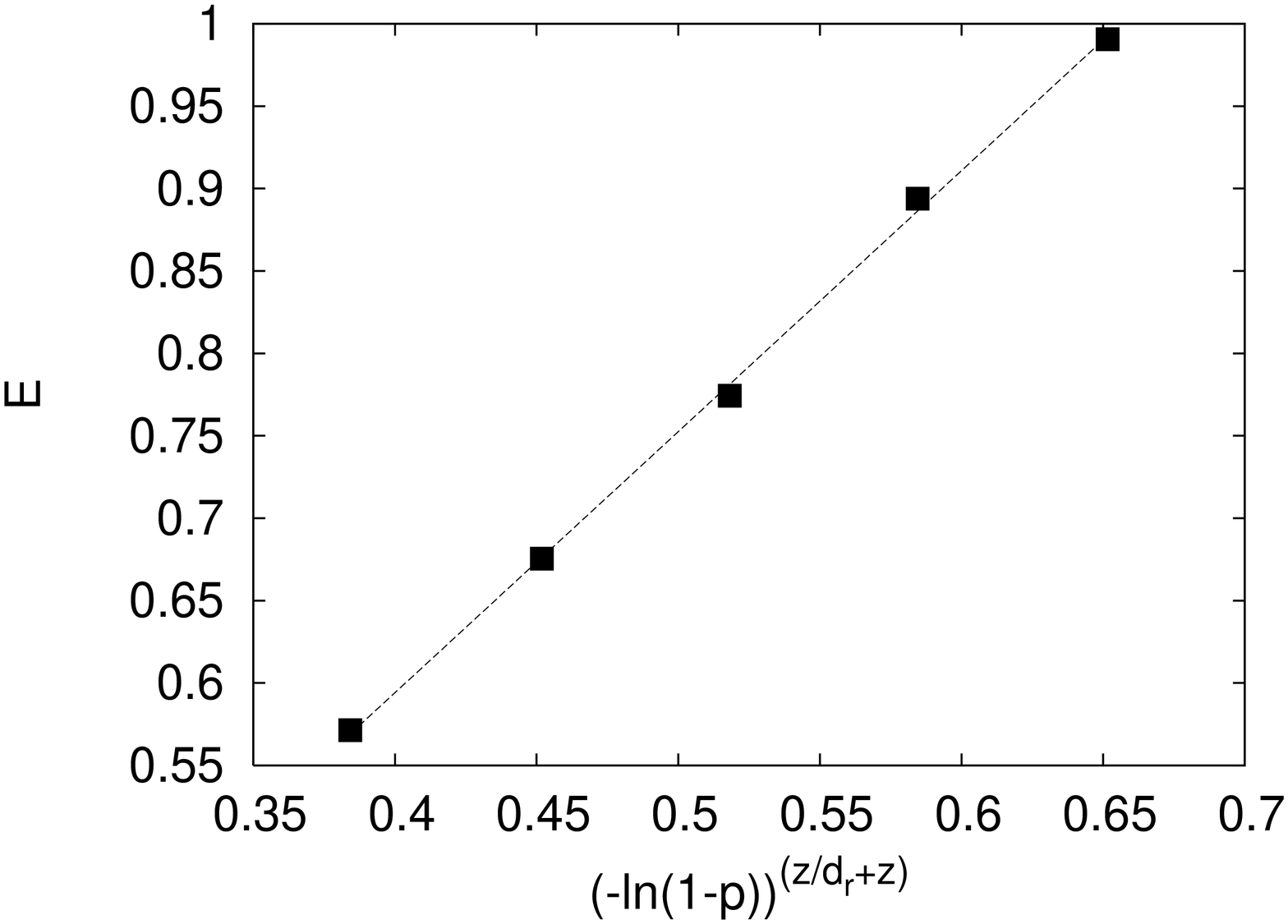}
\caption{ Left: Logarithm of the density at the clean critical point $\lambda_c^0$ as a
function of $t^{d_r/(d_r+z)}=t^{0.362}$ for several impurity concentrations
($p=0.2,\ldots,0.4$ from top to bottom) and $L=3000$. The long-time behavior follows a
stretched exponential $\ln \rho = -E t^{0.362}$. Right: Decay constant $E$ of the
stretched exponential as a function of $[-\ln(1-p)]^{z/(d_r+z)}=[-\ln(1-p)]^{0.638}$. }
\label{fig:stretched}
\end{figure}

According to eq.\ (\ref{eq:stretched}), the behavior right at the clean critical point,
$\lambda=\lambda_c^0$, is expected to be a stretched exponential rather than a simple
exponential decay. To shed more light on the time evolution at $\lambda_c^0$, the
behavior of $\ln \rho$ as a function of  $t^{d_r/(d_r+z)}$ is presented in the left panel
of figure \ref{fig:stretched} for several impurity concentrations $p$.
For our system, $d_r/(d_r+z)=0.362$ with $z=1.76$ being the dynamical exponent of the
clean 2d contact process \cite{voigt}. The figure shows that the data follow a stretched
exponential behavior $\ln \rho = -E t^{0.362}$ over more than three orders of magnitude
in $\rho$, in good agreement with eq. (\ref{eq:stretched}). (The very slight deviation of
the curves from a straight line can be attributed to the pre-exponential factors
neglected in the extremal statistics theory). The right panel of figure
\ref{fig:stretched} shows the decay constant $E$, i.e., the slope of these curves as a
function of $\tilde p =-\ln(1-p)$. In good approximation, the values follow the power law
$E \sim \tilde p^{z/(d_r+z)}=\tilde p^{0.638}$ predicted in (\ref{eq:stretched}).

\begin{figure}
\includegraphics[width=\figwidth,angle=-90]{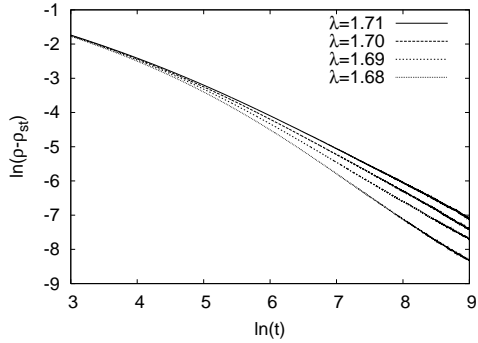}
\caption{Double-logarithmic plot of the approach of the density to its nonzero stationary
value in the tail of the smeared transition for a system with $p=0.2$ and $L=3000$ and
birth rate $\lambda=1.71, 1.70, 1.69, 1.68$ (top to bottom). The long-time behavior is of
power-law type, ($\rho(t)-\rho_{st}) \sim t^{-\psi}$. Fits yield exponents of
approximately 1.00, 1.08, 1.12, and 1.28, respectively.} \label{fig:powerlaw}
\end{figure}

In the tail of the smeared transition, i.e. for  $\lambda>\lambda_c^0$ the density has a
constant nonzero value $\rho_{st}=\rho(\infty)$ in the long-time limit. Figure
\ref{fig:powerlaw} illustrates the approach of the density to this value. It shows
$\ln[\rho(t)-\rho_{st}]$ as a function of $\ln(t)$ for several $\lambda$. The long-time
behavior is clearly of power-law type, but the exponent depends on $\lambda$, i.e., it is
nonuniversal. These results agree with the corresponding prediction in eq.\
(\ref{eq:power}).

\subsection{Stationary state}

\begin{figure}
\includegraphics[width=\figwidth,angle=-90]{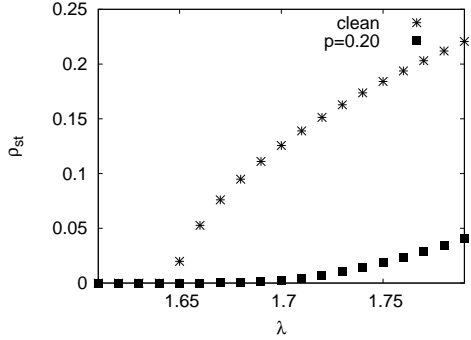}
\caption{Stationary density $\rho_{st}$ as a function of birth rate $\lambda$ for a clean
system and a system with impurity concentration $p=0.2$. System size is $L=1000$.}
\label{fig:density}
\end{figure}

In this subsection we present and analyze the simulation results for the stationary state
in the tail of the smeared transition,  $\lambda>\lambda_c^0$. Figure \ref{fig:density}
shows a comparison of the stationary density $\rho_{st}$ as a function of $\lambda$
between the clean system and a dirty system with $p=0.2$.
The clean system ($p=0$) has a sharp phase transition with a power-law singularity of the
density, $\rho_{st} \sim (\lambda-\lambda_c^0)^\beta$ with $\beta\approx 0.58$ in
agreement with the literature \cite{moreira}. In contrast, in the dirty system, the
density increases much more slowly with $\lambda$ after crossing the clean critical
point. This suggests either a critical point with a very large exponent $\beta$ or
exponential behavior.

\begin{figure}
\includegraphics[width=\figwidth,angle=-90]{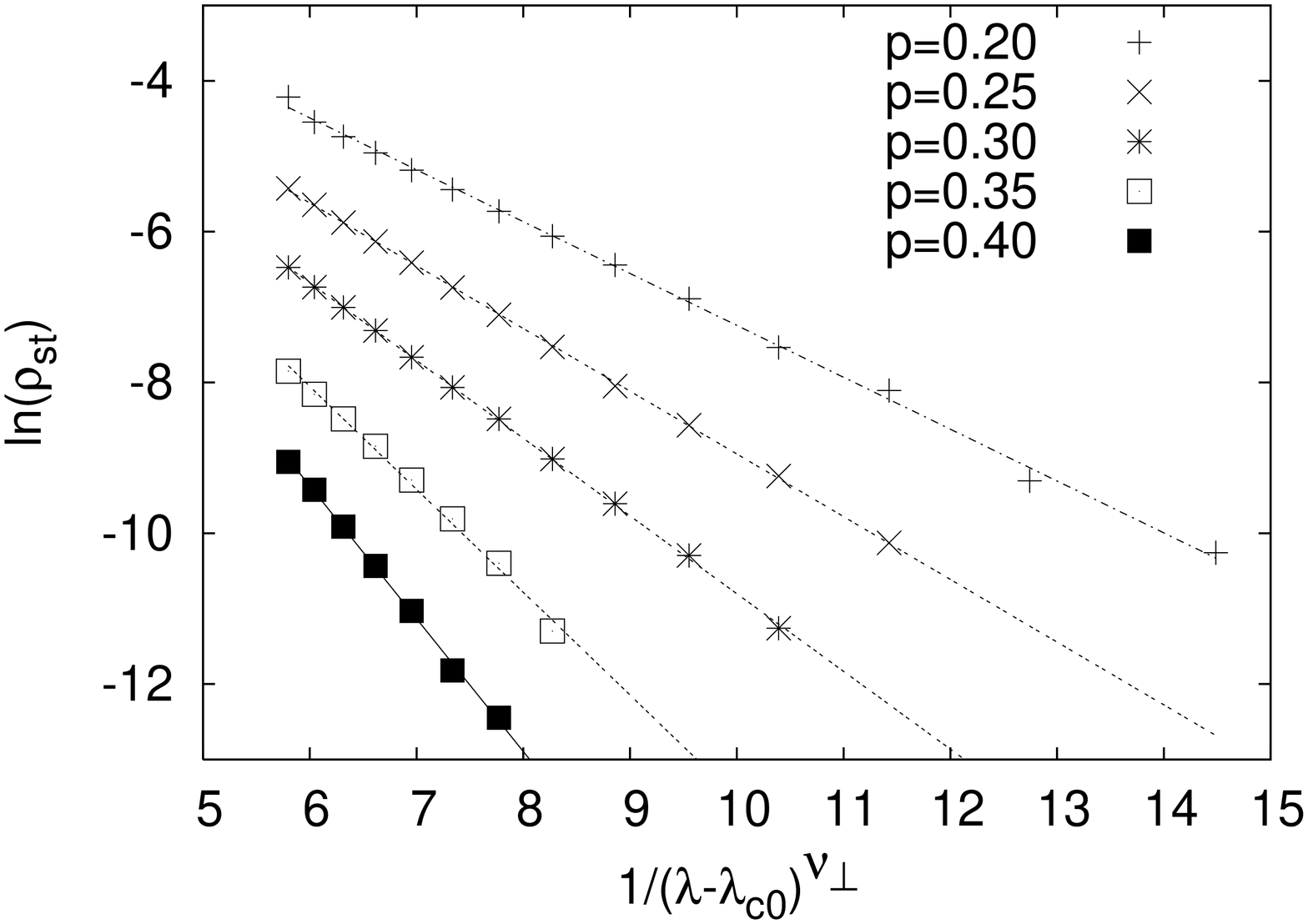}\includegraphics[width=\figwidth,angle=-90]{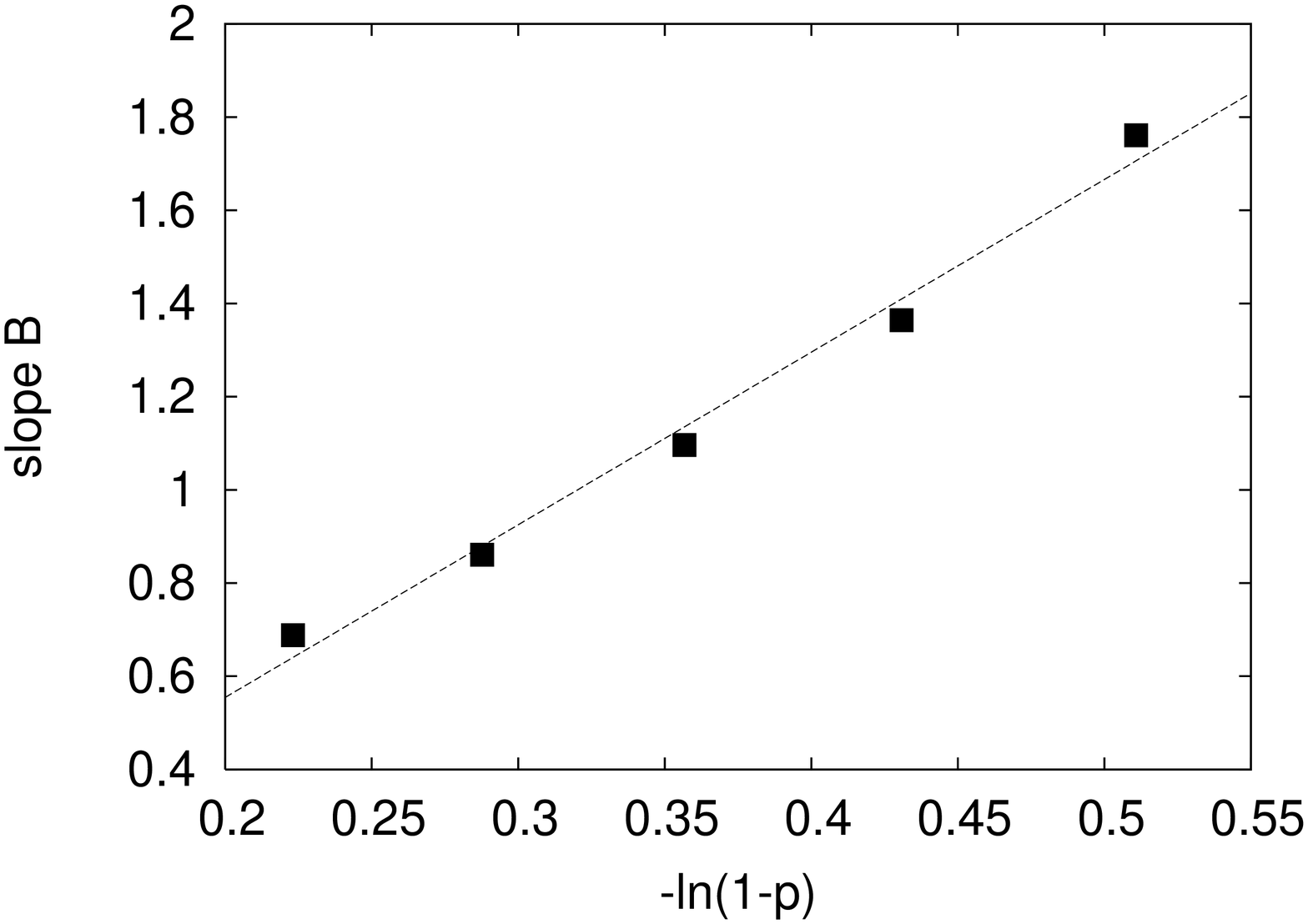}
\caption{Left: Logarithm of the stationary density $\rho_{st}$  as a function of
$(\lambda-\lambda_c^0)^{-d_r\nu_\bot}=(\lambda-\lambda_c^0)^{-0.734}$ for several
impurity concentrations $p$ and $L=3000$. The straight lines are fits to eq.\
(\ref{eq:rho}). Right: Decay constant $B$ as a function of $-\ln(1-p)$.} \label{fig:tail}
\end{figure}

Let us now investigate the behavior of the dirty system in the low-density tail more
closely.  In figure \ref{fig:tail}, we plot $\ln \rho_{st}$ as a function of
$(\lambda-\lambda_c^0)^{-d_r \nu_\bot}$ for several impurity concentrations $p$, as
suggested by eq.\ (\ref{eq:rho}).
The data in the left panel of figure \ref{fig:tail} show that the density tail is indeed
exponential, following the prediction $\ln \rho_{st} = -B (\lambda -\lambda_c^0)^{-d_r
\nu_\bot} $ over at least two orders of magnitude in $\rho_{st}$. (The clean 2d spatial
correlation length exponent is $\nu_\bot=0.734$ \cite{voigt}.)  Fits of the data to eq.\
(\ref{eq:rho}) are used to determine the decay constants $B$. The right panel of figure
\ref{fig:tail} shows these decays constants as function of $\tilde p = -\ln(1-p)$. The
dependence is close to linear, as predicted below eq.\ (\ref{eq:w_lam}) (Slight
deviations from the theoretical prediction can again be attributed to the pre-exponential
terms neglected in the extremal statistics theory.)

%%%%%%%%%%%%%%%%%%%%%%%%%%%%%%%%%%%%%%%%%%%%%%%%%%%%%%%%%%%%%%%%%%%%%%%%%%%%%%%%%%%%%%%%%%%%
\section{Discussion and conclusions}
\label{sec:conclusions}
%%%%%%%%%%%%%%%%%%%%%%%%%%%%%%%%%%%%%%%%%%%%%%%%%%%%%%%%%%%%%%%%%%%%%%%%%%%%%%%%%%%%%%%%%%%%

To summarize, we have provided extensive numerical evidence that extended impurities
destroy the sharp nonequilibrium phase transition in the contact process by smearing and
lead to a (nonuniversal) exponential dependence of the density and other quantities on
the control parameter. These results are in agreement with the predictions of Ref.\
\cite{contact_pre} which were based on extremal statistics arguments and mean-field
theory. In this section, we first relate our findings to a power-counting analysis of the
corresponding field theory. We then compare our smeared phase transition to the more
conventional Griffiths effects in the contact process with point-like
defects\cite{Griffiths,noest}, and we discuss general implications for theory and
experiment.

The field-theoretic formulation of the contact process \cite{Grassberger78,Cardy80} is
the so-called Reggeon field theory which was originally studied in the context of hadronic
interactions at ultra-relativistic energies (for a review see, e.g., \cite{Moshe}).
In appropriate units, its action reads
\begin{equation}
S=\int d^dr dt~ \tilde\phi \left[\partial_t -\kappa -\nabla^2+ \frac \Gamma 2
(\phi-\tilde\phi) \right]\phi
\end{equation}
where $\phi(\mathbf r,t)$ represents the density while $\tilde\phi(\mathbf r,t)$ is the
Martin-Siggia-Rosen response field and $\kappa$ denotes the bare distance from the
transition. A simple power counting analysis for the scale dimension of $\Gamma$ at the
Gaussian fixed point yields $[\Gamma]=4-d$, i.e., the upper critical dimension of this
field theory is $d_c^+=4$. Spatially quenched disorder (both uncorrelated or correlated)
can be taken into account by adding a term
\begin{equation}
S_{dis} = \gamma \int d^{d_r} r_r \left[ \int d^{d_{cor}} r_{cor}~ dt ~\tilde \phi \phi
\right]^2
\end{equation}
where the outer integral is over the uncorrelated directions $\mathbf r_r$ while the
inner spatial integrals are over the correlated directions $\mathbf r_{cor}$ (see
\cite{janssen97} for the uncorrelated case $d_{cor}=0$). Power counting for the scale
dimension of $\gamma$ gives $[\gamma] =4-d+d_{cor}$. Uncorrelated disorder is marginal at
$d=d_c^+=4$, but disorder correlations increase the scale dimension of $\gamma$ making
the disorder term renormalization-group relevant. The power-counting analysis thus
predicts stronger effects for correlated disorder than for point-like disorder, in
agreement with our results. Let us emphasize however, that the smeared phase transition
scenario found in the present manuscript cannot be obtained in a perturbative analysis of
the Reggeon field theory because the rare regions are non-perturbative degrees of
freedom.

Both conventional Griffiths effects and the smearing scenario found in the present paper
are caused by rare large spatial regions which are locally in
the active phase even if the bulk system is not. The difference between Griffiths effects
and the smearing of the transition is the result of disorder correlations. For point-like
defects, i.e., uncorrelated disorder, the rare regions are of finite size and cannot
undergo a true phase transition. Instead, they fluctuate slowly which gives rise to
Griffiths effects. In contrast, if the rare regions are infinite in at least one
dimension, a stronger effect occurs: each rare region can independently undergo the phase
transition and develop a nonzero steady state density. This leads to a smearing of the
global transition.

The smearing mechanism found here relies only on the existence of a true phase transition
on an isolated rare region. It should therefore apply not only to the directed
percolation universality class, but to an entire family of nonequilibrium universality
classes for spreading processes and reaction-diffusion systems. Note that while the
presence or absence of smearing is universal in the sense of critical phenomena (it
depends on symmetries and dimensionality only), the functional form of the density and
other observables is {\em not} universal, it depends on the details of the disorder
distribution \cite{us_planar}.

Smearing phenomena similar to the one found here can also occur at equilibrium phase
transitions. At quantum phase transitions in itinerant electron systems, even point-like
impurities can lead to smearing \cite{us_rounding} (the necessary disorder correlations
are in imaginary time direction). In contrast, for the classical Ising (Heisenberg)
universality class, the impurities have to be at least 2d (3d) for the transition to be
smeared which makes the phenomenon less likely to be observed \cite{us_planar}.

In the context of our findings it is worth noting that, despite its ubiquity in theory
and simulations, clearcut experimental realizations of the directed percolation
universality class are strangely lacking \cite{hinrichsen_exp}. To the best of our
knowledge, the only verification so far has been found in the spatio-temporal
intermittency in ferrofluidic spikes \cite{spikes}.  We suggest that the disorder-induced
smearing found in the present paper may explain the striking absence of directed
percolation scaling \cite{hinrichsen_exp} in at least some of the experiments.

\ack We thank R. Sknepnek and U. T\"auber for stimulating discussions. We acknowledge
support from the NSF under grant No. DMR-0339147. Part of this work has been performed at
the Aspen Center for Physics.



\section*{References}
\begin{thebibliography}{99}
\frenchspacing
\bibitem{Griffiths} Griffiths R B 1969  {\it \prl} {\bf 23} 17
\bibitem{dsf9295} Fisher D S 1995 {\it \prb} {\bf 51} 6411
\bibitem{McCoyWu} McCoy B M and Wu T T 1968 {\it \prl} {\bf 23} 383 (1968)
    \par\item[]   McCoy B M and Wu T T 1968 {\it Phys. Rev.}  {\bf 176} 631
\bibitem{us_planar} Vojta T 2003 {\it J. Phys. A} {\bf 36} 10921
\bibitem{us_rounding} Vojta T 2003 {\it \prl} {\bf 90} 107202
\bibitem{chopard_book} Chopard B and Droz M 1998 {\it Cellular
    Automaton Modeling  of Physical Systems} (Cambridge University Press,
    Cambridge, England).
\bibitem{marro_book} Marro J and Dickman R 1999 {Nonequilibrium Phase Transitions in
    Lattice Models} (Cambridge University Press,
    Cambridge, England).
\bibitem{hinrichsen00} Hinrichsen H 2000 {\it Adv. Phys.} {\bf 49} 815
\bibitem{tauber} T\"auber U C 2003 {\it Adv. in Solid State Phys.} {\bf 43} 659
\bibitem{dp} Grassberger P and  de la Torre A 1979 {\it Ann. Phys. (NY)} {\bf 122} 373
\bibitem{conjecture} Janssen H K 1981  {\it Z. Phys. B} {\bf 42} 151
    \par\item[]    Grassberger P 1982 {\it Z. Phys. B} {\bf 47} 365
\bibitem{contact} Harris T E 1974 {\it Ann. Prob.} {\bf 2} 969
\bibitem{ziff} Ziff R M, Gulari E, and Barshad Y 1986 {\it \prl} {\bf 56} 2553
\bibitem{tang} Tang L H and Leschhorn H 1992 {\it \pra} {\bf 45} R8309
\bibitem{turb} Pomeau Y {\it Physica D} {\bf 23} 3
\bibitem{harris} Harris A B 1974 {\it J.\ Phys. C} {\bf 7} 1671
\bibitem{noest} Noest A J 1986 {\it \prl} {\bf 57} 90
\bibitem{janssen97} Janssen H K 1997 {\it \pre} {\bf 55} 6253
\bibitem{bramson} Bramson B, Durrett R  Schonmann R H 1991 {\it Ann. Prob.} {\bf 19} 960
\bibitem{moreira} Moreira A G and Dickman R 1996 {\it \pre} {\bf 54} R3090
\bibitem{webman} Webman T et al. 1998 {\it Phil. Mag. B} {\bf 77} 1401
\bibitem{cafiero} Cafiero R,  Gabrielli A and Mu\~{n}oz M A 1998 {\it \pre} {\bf 57} 5060
\bibitem{alcaraz} Alcaraz F C 1994 {\it Ann. Phys. (NY)} {\bf 230} 250
\bibitem{hooyberghs} Hooyberghs J, Igloi F and Vanderzande C 2003 {\it \prl} {\bf 90} 100601
    \par\item[] Hooyberghs J, Igloi F and Vanderzande C 2004 eprint cond-mat/0402086
\bibitem{SDRG} Ma S K, Dasgupta C and  Hu C-K 1979 {\it \prl} {\bf 43} 1434
\bibitem{contact_pre} Vojta T 2004 {\it \pre} {\bf 70} 026108
\bibitem{barber} Barber M N 1983 in {\it Phase transitions and critical phenomena}, edited by C.
     Domb and J.L. Lebowitz (Academic, London), Vol. 8.
\bibitem{dickman99} Dickman R 1999 {\it \pre} {\bf 60} R2441
\bibitem{voigt} Voigt C A and Ziff R M 1997 {\it \pre} {\bf 56} R6241
\bibitem{Grassberger78} Grassberger P and Sundermeyer K 1978 {\it Phys. Lett. B} {\bf 77} 220
\bibitem{Cardy80} Cardy J L and Sugar R L 1980 {\it J. Phys A} {\bf 13} L423
\bibitem{Moshe} Moshe M 1978 {\it Phys Rep C} {\bf 37} 257
\bibitem{hinrichsen_exp} Hinrichsen H 2000  {\it Braz. J. Phys.} {\bf 30} 69
\bibitem{spikes} Rupp P, Richter R and Rehberg I {\it \pre} {\bf 67} 036209

\end{thebibliography}
\end{document}